\documentclass[11pt, aps, preprint, nofootinbib,superscriptaddress,eqsecnum,titlepage]{revtex4-2}
\usepackage[active]{srcltx}
\usepackage[utf8]{inputenc}
\usepackage{latexsym}
\usepackage{hyperref}
\usepackage{amsmath}
\usepackage{amssymb}
\usepackage{amsfonts}
\usepackage{graphicx}
\usepackage{slashed}
\usepackage{xcolor}
\usepackage{soul}
\usepackage{slashed}
\usepackage{braket}
\usepackage{natbib}
\usepackage[export]{adjustbox}
\usepackage{relsize}
\newcommand{\xdownarrow}[1]{%
	{\left\downarrow\vbox to #1{}\right.\kern-\nulldelimiterspace}
}

\usepackage{MnSymbol}
\usepackage{wasysym}

\makeatletter
\newcommand*{\sumcirclearrowleft}{%
  \DOTSB
  \mathop{
    \mathchoice
      {\rlap{\kern.25em\rotatebox[origin=c]{-90}{$\circlearrowleft$}}{\sum}}
      {\vcenter{\rlap{\kern.2em\rotatebox[origin=c]{-90}{$\scriptscriptstyle\circlearrowleft$}}}{\sum}}
      {\sum}{\sum}
  }\slimits@
}

\begin{document}

\title{\Large{{\sc Quantum wires, Chern-Simons theory, and dualities in the quantum Hall system}}}

\author{Julio Toledo}
\email{juliotoledo@uel.br}
\affiliation{Departamento de Física, Universidade Estadual de Londrina,  86057-970, Londrina, PR, Brasil}

\author{Renann Lipinski Jusinskas}
\email{renannlj@fzu.cz}
\affiliation{Institute of Physics of the Czech Academy of Sciences \& CEICO, Na Slovance 2, 18221 Prague, Czech Republic}

\author{Carlos A. Hernaski}
\email{carloshernaski@utfpr.edu.br}
\affiliation{Departamento de Física, Universidade Tecnológica Federal do Paraná, 85503-390, Pato Branco, PR, Brasil}

\author{Pedro R. S. Gomes}
\email{pedrogomes@uel.br}
\affiliation{Departamento de Física, Universidade Estadual de Londrina,  86057-970, Londrina, PR, Brasil}


\begin{abstract}

Over the years, many theoretical frameworks have been developed to understand the remarkable physics of the quantum Hall system. In this work we discuss the interplay among quantum wires, Chern-Simons theory, bosonization, and particle-vortex duality, which enable a unified approach for the Laughlin states of the fractional quantum Hall system. Starting from the so-called quantum wires system, which is a semi-microscopic description in terms of 1+1 dimensional theories, we discuss the emergence of 2+1 dimensional low-energy effective field theories by using different maps connecting the microscopic degrees of freedom with the macroscopic ones. We show the origin of these maps by embedding the bosonized version of the original quantum wires model in a gauge invariant theory and using particle vortex-duality.

\end{abstract}

\maketitle
\tableofcontents


\section{Introduction}

The Quantum Hall Effect (QHE) is one of the crown jewels of condensed matter physics. While its conception is simple, the underlying  physics is extremely rich, with far-reaching implications that  have ultimately driven the development of the whole field of topological phases of matter \cite{Wen:1992vi, Wen:1995qn, Wen:2012hm}. Theoretical descriptions of the QHE go back to Laughlin \cite{laughlin1983anomalous}, and many more have been proposed since then to model and understand this formidable phenomenon \cite{Girvin:1987quantum, Jain:2007composite,Tong:2016kpv}. The present work aims to precisely connect two contrasting descriptions of the Laughlin series of the fractional QHE, either using effective field theories (EFTs) or the so-called quantum wires (QW) approach.

The EFTs we focus on are 2+1 dimensional field theories with dual descriptions of the quasi-particle excitations. On one hand, the quasi-particles are seen as elementary excitations of some field, and the respective EFTs are usually referred to as Landau-Ginzburg-Chern-Simons or Zhang-Hansson-Kivelson (ZHK) theories \cite{zhang1989effective,Zhang:1992eu}. On the other hand, the EFTs describing quasi-particles excitations in terms of vortices are known as Wen-Zee theories or hydrodynamical topological field theories \cite{Wen:1995qn,wen1990compressibility}.

The quantum wires approach establishes a partially microscopic description of the QHE, with interactions modelled directly in terms of electron degrees of freedom propagating in an array of one-dimensional wires. This approach was pioneered by \cite{kane2002fractional,teo2014luttinger} in the description of the  fractional QHE and since then has been a subject of great interest, being used recently in several investigations of topologically ordered systems. Indeed, it has been generalized to other Abelian and non-Abelian quantum Hall phases \cite{Klinovaja:2014ifqe,Klinovaja:2014iba,Klinovaja:2014ahe,Cano:2014pya,Fuji:2016lgv,Fuji:2018fuj,Kane:2018amn}, used in the description of topological insulators \cite{Neupert:2014nga,Sagi:2014yea,santos2015fractional,Santos:2019orbit} and superconductors \cite{Sahoo:2015zfa, Park:2018fwb}, in the constructions of topologically ordered spin liquids \cite{Gorohovsky:2015fya, Meng:2015gya, Huang:2016jms, Patel:2016jwi, Hernaski:2017agt}, and also extended to higher dimensional topological phases \cite{Meng:2015fractop, Sagi:2015zea, iadecola2016wire}. For a recent review, see \cite{Meng:2019ket}.

A natural step towards a better understanding of the QHE is to investigate whether the different descriptions available share an underlying connection, leading to a more unified theoretical framework. While this is still far from being achieved for general fractional quantum Hall phases, a more favorable scenario emerges  in the class of Laughlin states of the Abelian QHE \cite{laughlin1983anomalous}, characterized by the filling fraction of the form $\nu=1/m$, with $m$ being an odd integer. In particular, explicit connections between QW and EFT descriptions have been obtained in \cite{santos2015fractional,fontana2019quantum,imamura2019coupled}. In \cite{fontana2019quantum} and \cite{imamura2019coupled}, however, the maps relating the microscopic degrees of freedom and the fields of the effective theory are rather different, and their equivalence was unclear\footnote{Perhaps for this reason the authors of \cite{imamura2019coupled} did not recognize a bigger picture connecting our works.}. One of the main purposes of this work is to fill this gap, showing that the approaches of \cite{fontana2019quantum} and \cite{imamura2019coupled} are indeed equivalent. The pivot underpinning all these connections is the particle-vortex duality, with explicit realizations in terms of quantum wires \cite{Mross:2015idy, mross2017symmetry}.

In one of its simplest incarnations, the particle-vortex duality is the equivalence between the two 2+1 dimensional relativistic theories,
\begin{eqnarray}
	\label{bosPV}
S_1 \equiv \int |D_B\phi|^2-V(\phi)\quad \Leftrightarrow \quad S_2\equiv \int  |D_\alpha\tilde\phi|^2-\tilde{V}(\tilde\phi)+\dfrac{1}{2\pi}Bd\alpha,\label{1}
\end{eqnarray}
where $B$ is a background gauge field for the global $U(1)$ symmetry and $\alpha$ is a dynamical gauge field, with covariant derivatives $D_B\equiv \partial-iB$ and $D_{\alpha}\equiv \partial-i\alpha$. In addition, the potentials $V(\phi)$ and $\tilde{V}(\tilde{\phi})$ admit spontaneous symmetry breaking, e.g.,
\begin{equation}
V(\phi)=M^2|\phi|^2+\frac{\lambda}{4}|\phi|^4~~~\text{and}~~~\tilde{V}(\tilde{\phi})=\tilde{M}^2|\tilde{\phi}|^2+\frac{\tilde{\lambda}}{4}|\tilde{\phi}|^4.\label{2}
\end{equation}
The background field $B$ couples to the particle current in the action $S_1$ and to the topological (vortex) current in the action $S_2$. This is a salient feature of the particle-vortex duality. The matching of the phases is achieved through the correspondence $M^2 \Leftrightarrow -\tilde{M}^2$. Thus, the gapped phase $M^2>0$ of the theory $S_1$ corresponds to the Higgs phase $\tilde{M}^2<0$ of $S_2$, whereas the Goldstone bosons of $S_1$ correspond to the photon of the theory $S_2$ in the gapless phase.

The particle-vortex duality has been recently found to be a central element in the so-called {\it web of dualities}, a remarkable  series of connections between 2+1 dimensional quantum field theories \cite{karch2016particle,seiberg2016duality, Murugan:2016zal}. The web of dualities started with Son's groundbreaking work in \cite{son2015composite}, who suggested a fermionic counterpart to the particle-vortex duality in the description of the QHE at the metallic filling fraction $\nu=1/2$. Enlightening reviews can be found in \cite{senthil2019duality, Turner:2019wnh}. In particular, the quantum wires formulation provides a fruitful perspective on bosonic and fermionic particle-vortex dualities. It leads to explicit mappings between dual theories through bosonization \cite{Mross:2015idy,mross2017symmetry}, placing the dualities on a sounder basis. 

In the context of QHE and building on the results of \cite{mross2017symmetry}, Fuji and Furusaki were able to cast the quantum wires description of the Abelian fractional QHE in different forms by using certain maps between microscopic degrees of freedom \cite{Fuji:2018fuj}. While it is possible to extract the physics of the different microscopic formulations and match the corresponding ZHK and Wen-Zee effective theories, explicit connections between microscopic degrees of freedom and fields in the continuum are very difficult to be established.  In the infrared (IR) limit, however, it is possible to find explicit maps, as presented in  \cite{fontana2019quantum} and \cite{imamura2019coupled}.

In this work we further explore these ideas, with the main goal of showing that the approaches of \cite{fontana2019quantum} and \cite{imamura2019coupled} connecting quantum wires with low-energy effective theories are equivalent, and how their equivalence follows from the particle-vortex duality.
With this purpose, we start by discussing that the ZHK and Wen-Zee effective field theories are related by a particle-vortex duality.
Both ZHK and Wen-Zee effective theories involve emergent gauge structures, namely a statistical gauge field in the case of ZHK theory, responsible for attaching flux to the bosons to turn then into fermions, and a hydrodynamical gauge field in the case of Wen-Zee theory, capturing the collective behavior of the electrons in the QHE.  We then analyze the fate of these structures in the IR limit. In ZHK theory the gauge structure is manifested in the broken (Higgs) phase, whereas in Wen-Zee theories the gauge structure appears in the symmetric phase. 
In the infrared (IR) limit these theories flow to Self-Dual (SD) and to Maxwell-Chern-Simons (MCS) theories, respectively. 
In other words, we reinterpret the long known relationship between SD and MCS theories \cite{deser1984self} as a particle-vortex duality. This is the first hint of how the works of \cite{fontana2019quantum} and \cite{imamura2019coupled} are related, once they connect the quantum wires directly with MCS and SD theories, respectively.

We then describe the Laughlin series of the fractional QHE in terms of quantum wires and embed the model in a gauge invariant theory. 
This leads to a direct identification with the ZHK model in its IR limit, which in the unitary gauge corresponds to the SD model. We then apply particle-vortex transformations on the quantum wires variables and show how the Wen-Zee theory emerges in this scenario, as well as its IR limit in terms of the MCS model. The duality between quantum wires theories and their low-energy effective actions can also be explored in terms of particle and vortex creation operators. We also show that the vortex creation operators in the dual quantum wires models are mapped to the respective vortex creation operators in the low-energy effective gauge theories. In a different perspective, we interpret the gauge field prescriptions in terms of quantum wires variables of work \cite{fontana2019quantum} as a momentum-winding duality between ZHK and MCS, namely, a generalization of the known duality between the XY model and Maxwell theory in 2+1 dimensions. In sum, this work provides a unified framework for describing the Laughlin states of the quantum Hall effect.

This paper is organized as follows. In Sec. \ref{macroscopicsec}, we discuss the relevant aspects of the ZHK and Wen-Zee theories, their infrared limit, and their Hamiltonian formulation, which is useful for the comparison with the quantum wires. In Sec. \ref{quantumwiressec}, we review the quantum wires description of the Laughlin states and show the explicit connections with the continuum theories. Sections \ref{macroscopicsec} and \ref{quantumwiressec} are mostly of review character but nevertheless offer new perspectives on previously known facts and also contain some new presentations that lead to the main results of the work, contained in Sec. \ref{eft}. In Subsections \ref{123} and \ref{MCS}, we embed the quantum wires in a gauge invariant model and use particle-vortex duality on the wires lattice to justify the prescriptions of \cite{fontana2019quantum} and \cite{imamura2019coupled} that lead to the MCS and SD models in the IR regime, respectively.  In Sec. \ref{mw} we generalize the XY-Maxwell duality and interpret the prescription of \cite{fontana2019quantum} as a momentum-winding duality between ZHK and MCS models. In Sec. \ref{vo}, we construct the vortex creation operators in the dual quantum wires models and show that they are mapped through the gauge field prescriptions to the corresponding vortex creation operators in the low-energy gauge theories. We conclude in Sec. \ref{discussionsec} with a brief summary and a discussion of the results and possible implications. Some details of the computations involved here are presented in the Appendix \ref{AA}.


\section{Effective Field Theories for Laughlin States}\label{macroscopicsec}

In this section we discuss the ZHK and Wen-Zee theories for Laughlin states, relating them through the particle-vortex duality. We then consider their infrared (IR) limit and corresponding Hamiltonian formulations, which are useful in the comparison with the quantum wires approach in later sections.

\subsection{Particle-Vortex Duality}

Effective field theory descriptions of the Laughlin class of Abelian fractional QHE can be embedded into the bosonic particle-vortex duality \cite{1991IJMPB...5.2675L,Burgess:2000kj,Zee:1995avy,Tong:2016kpv}. The basic idea is to gauge the global $U(1)$ symmetry of \eqref{bosPV} to make the gapless excitations unphysical, along with the introduction of suitable Chern-Simons terms.  More explicitly, we introduce in both sides of \eqref{bosPV} the term $\dfrac{k}{4\pi}BdB+\dfrac{1}{2\pi}AdB$, and promote $B$ to a dynamical gauge field $a$, with $A$ denoting the external electromagnetic field used to measure Hall conductivity\footnote{We use the notation that capital letters $A,B,...$ denote external fields, whereas lowercase letters $a,b,\alpha,...$ represent dynamical ones.}.

This sequence of operations leads to the relation
\begin{eqnarray}
	S_1&=&\int  |D_a\phi|^2-V(\phi)+\dfrac{k}{4\pi}ada +\dfrac{1}{2\pi}Ada \nonumber\\ \Leftrightarrow\quad  S_2&=&\int   |D_\alpha\tilde\phi|^2-\tilde{V}(\tilde\phi)+\dfrac{1}{2\pi}ad(\alpha+A)+\dfrac{k}{4\pi}ada.
\label{deformedPVwithsources}		
\end{eqnarray}
We can further reduce the theory $S_2$ by making the shift $\alpha\rightarrow \alpha-A$, and integrating out the field $a$ that appears quadratically, 
\begin{eqnarray}
S_1&=&\int  |D_a\phi|^2-V(\phi)+\dfrac{k}{4\pi}ada +\dfrac{1}{2\pi}Ada\nonumber\\ \Leftrightarrow\quad S_2&=& \int  |D_{\alpha-A}\tilde\phi|^2-\tilde{V}(\tilde\phi)-\dfrac{1}{4\pi k}\alpha d\alpha.
\label{deformedPVwithbackground}
\end{eqnarray}
The action $S_1$ is the Wen-Zee effective description of the Laughlin class of the fractional QHE. The electromagnetic field is coupled to the topological current $J=\frac{1}{2\pi}da$, parametrized by the hydrodynamic field $a$. The action includes  a Chern-Simons term for the field $a$ with a properly quantized level. This emergent field couples to the quasi-particle current, written in terms of the scalar field $\phi$. Electron excitations correspond to the vortices of $\phi$. The action $S_2$ is the relativistic counterpart of the original ZHK description of QHE (the nonrelativistic setting will be discussed below). The external field couples to the particle current of the field $\tilde{\phi}$. The statistical field $\alpha$, without a properly quantized Chern-Simons level, attaches an odd number $m$ of flux units to the $\tilde\phi$-excitations, which effectively turns them into fermions. This is  known as the composite boson picture. In this case, quasi-particles are described by the vortices of the field $\tilde{\phi}$. 

It is useful for our purposes to consider the IR limit of the duality \eqref{deformedPVwithbackground} in the phase describing the Laughlin state. It corresponds to the symmetric phase of $S_1$ ($M^2>0$) and to the Higgs phase of $S_2$ ($\tilde{M}^2<0$). Then, we have 
\begin{equation}
	\int -\dfrac{1}{96\pi M}f_a^2+\dfrac{k}{4\pi}ada+\dfrac{1}{2\pi}Ada \quad \Leftrightarrow\quad \int \dfrac{|\tilde{M}^2|}{2\tilde\lambda}(\alpha_\mu-A_\mu)(\alpha^\mu-A^\mu)-\dfrac{1}{4\pi k}\alpha d\alpha,
\label{nontrivialphases}
\end{equation}
where the Maxwell term in theory $S_1$ comes from one-loop contributions. There are several interesting points in this relation. First, we observe in \eqref{deformedPVwithbackground} that the scalar field $\tilde{\phi}$ is charged under both $\alpha$ and $A$. In the Higgs phase \eqref{nontrivialphases}, we have absorbed the phase of the field into $\alpha_{\mu}$, which means that the redefined $\alpha_{\mu}$ field is now charged under the $U(1)_A$. Second, in the deep IR limit ($M$ and $|\tilde{M}^2|\rightarrow\infty$), integrating out the dynamical fields reduces both sides of \eqref{nontrivialphases}  to $-\frac{1}{4\pi k}AdA$, yielding the Hall conductivity of the Laughlin state $\sigma_{xy}=\frac{1}{2\pi k}$. Finally, by turning off the external field $A$ in \eqref{nontrivialphases}, we obtain the known duality between Maxwell-Chern-Simons (MCS) and the Self-Dual (SD) Model  \cite{deser1984self}, leading to the identification
\begin{equation}
	|\tilde{M}^2|=\dfrac{6\tilde\lambda}{\pi }M.
\label{eq:macroscopicrelationofMs}
\end{equation}
Therefore, the MCS-SD duality by itself can also be framed as a particle-vortex type. This is supported simply by the fact that the electromagnetic field couples to the vortex current $\frac{1}{2\pi}da$ in MCS, whereas it couples to the (``Higgsed'') particle current $-\frac{|\tilde{M}^2|}{\tilde\lambda}(\alpha-A)$ in the SD model in  \eqref{nontrivialphases}. The form of the duality in \eqref{nontrivialphases} will be important later when connecting the quantum wires description with field theories in the continuum limit.

\subsection{Nonrelativistic Setting}

Next, it will be enlightening to discuss the nonrelativistic counterpart of the duality \eqref{deformedPVwithbackground} \cite{1991IJMPB...5.2675L,Zee:1995avy}. We simply replace $\phi$ and $\tilde{\phi}$, that create both particle and anti-particles, by their nonrelativistic counterparts $\psi$ and $\tilde{\psi}$, accompanied by the respective chemical potentials $\mu$ and $\tilde{\mu}$ to control the average number of particles. The nonrelativistic duality then reads
\begin{eqnarray}
\label{nr_duality}
&&S_1=\int i\psi^*(\partial_t - i a_t- i\mu)\psi -\dfrac{1}{2M}|(\partial_i-i a_i)\psi|^2-V(\psi)+\dfrac{k}{4\pi}ada+\dfrac{1}{2\pi}Ada\\  \Leftrightarrow ~ && S_2=\int i\tilde\psi^*(\partial_t - i (\alpha_t-A_t)- i\tilde{\mu})\tilde\psi -\dfrac{1}{2\tilde{M}}|(\partial_i-i(\alpha_i-A_i))\tilde\psi|^2-V(\tilde{\psi})-\dfrac{1}{4\pi k}\alpha d \alpha.\nonumber
\end{eqnarray}

As before, we consider the IR limit describing the Laughlin state in the nonrelativistic case. Relying only on gauge invariance, now without Lorentz covariance, the integration over the matter fields produces
\begin{eqnarray}\
	\int {c}_1 f_{a,0i}^2+{c}_2f_{a,ij}^2 +\dfrac{k}{4\pi} ada +\dfrac{1}{2\pi}Ada ~ \Leftrightarrow ~ \int \tilde{c_1}( \alpha_0-A_0)^2+\tilde{c_2}(\vec{\alpha}-\vec{A})^2 -\dfrac{1}{4\pi k}\alpha d\alpha,
\label{NRGaugeduality}
\end{eqnarray}
where $c_1\neq c_2$ and $\tilde{c}_1\neq\tilde{c}_2$ are constants depending on the parameters of the theories in \eqref{nr_duality}. The above relation is the nonrelativistic version of the duality between Maxwell-Chern-Simons and the Self-Dual model, which can be established with the usual construction of the interpolating (master) Lagrangian. As a by-product of this analysis, we will be able to find the relation between the  parameters $c_1,c_2$ and $\tilde{c}_1, \tilde{c}_2$.  The master Lagrangian is
\begin{eqnarray}
	\label{masterlagr}
	\mathcal{L}_{master}[a,\alpha;A] &=& q_1(\alpha_0-A_0)^2+q_2(\alpha_i-A_i)^2+\dfrac{1}{2\pi}\alpha da+\dfrac{k}{4\pi}ada,
\end{eqnarray}
with $q_1, q_2, q_3$ denoting constant parameters. As we will see, integrating out the fields $(\alpha_0,\alpha_i)$ or $(a_0,a_i)$ leads to MCS or SD theory, respectively. 

Let us start with the integration of $(a_0,a_i)$. Their equations of motion can be cast as
\begin{eqnarray}
	d\alpha +kda=0\quad \Rightarrow\quad a = -\dfrac{1}{k}\alpha.\label{10}
\end{eqnarray}
Plugging them back in the master Lagrangian leads to
\begin{eqnarray}
	\mathcal{L}_{master}[\alpha;A] 	=  q_1(\alpha_0-A_0)^2+q_2(\alpha_i-A_i)^2-\dfrac{1}{4\pi k}\alpha d\alpha.\label{11}
\end{eqnarray}
Now, comparing this Lagrangian with the Self-Dual model in \eqref{NRGaugeduality}, we readily see that
\begin{eqnarray}
	q_1 = \tilde{c}_1 \quad\text{and} \quad q_2=\tilde{c}_2.
	\label{r1}
\end{eqnarray}

On the other hand, the equations of motion of $\alpha$ can be expressed as
\begin{eqnarray}
\alpha_0 =A_0 -\dfrac{1}{4\pi q_1}\epsilon_{ij}\partial_i a_j,\quad \alpha_i =A_i+\dfrac{1}{4 \pi q_2}\epsilon_{ij}f_{a,0j},\label{13}
\end{eqnarray}
such that the master Lagrangian is rewritten as
\begin{eqnarray}
\mathcal{L}_{master}[a;A]=-\dfrac{1}{16\pi ^2q_2}f_{a,0i}^2 -\dfrac{1}{ 32\pi^2 q_1}f_{a,ij}^2 + \dfrac{k}{4\pi}ada+\dfrac{1}{2\pi} Ada.\label{14}
\end{eqnarray}
A direct comparisson with the Maxwell-Chern-Simons theory \eqref{NRGaugeduality} yields
\begin{eqnarray}
 -\dfrac{1}{16\pi^2q_2} = c_1	\quad\text{and}\quad-\dfrac{1}{32\pi^2q_1}=c_2, 
 \label{r2}
\end{eqnarray}
Finally, combining the identifications \eqref{r1} and \eqref{r2} leads to
\begin{eqnarray}
c_1\tilde{c}_2 =-\frac{1}{16\pi^2} \quad\text{and}\quad  c_2\tilde{c}_1=-\frac{1}{32\pi^2}.\label{16}
\end{eqnarray}
In the relativistic case ($c_1=-2c_2$ and $\tilde{c}_1=-\tilde{c}_2$) these two equations reduce to a single one.



\subsection{Hamiltonian Analysis}\label{subsec:hamiltoniananalysis}

Now we would like to express the dual theories of \eqref{nontrivialphases} in a form that can be directly compared with the quantum wires formulation. In this case, the Hamiltonian description is more transparent.

The canonical momenta in the MCS theory \eqref{nontrivialphases} are
\begin{equation}
\Pi_0=0~~~\text{and}~~~\Pi_i=\frac{1}{24\pi M} f_{0i}+\frac{1}{2\pi}\epsilon_{ij}\left(A_j+\frac{k}{2}a_j\right).\label{17}
\end{equation}
The first relation defines a primary constraint. Then, the canonical Hamiltonian is given by
\begin{eqnarray}
\mathcal{H}_{MCS}&=&\Pi_i (\partial_0 a_i)-\mathcal{L}\nonumber\\
&=&\frac{1}{48\pi M} \left(\vec{e}^2+b^2\right) - \frac{a_0}{2\pi}\left(B+k\,b + \frac{1}{12M} \nabla\cdot \vec{e}\right) -\frac{1}{2\pi} A_0 b. \label{18}
\end{eqnarray}
Here we have introduced the electric and magnetic fields $e_i\equiv f_{0i}$ and $b\equiv \epsilon_{ij}\partial_{i}a_j$, respectively. The component $a_0$ plays the role of a Lagrange multiplier enforcing the Gauss's law
\begin{equation}
B+k \,b +\frac{1}{12 M} \nabla\cdot \vec{e}=0,
\label{eq:constraintMCS}
\end{equation}
which gives a secundary constraint and ensures the time independence of the primary constraint $\Pi_0=0$.
On the constrained surface defined by the full set of constraints, the Hamiltonian reduces to  
\begin{equation}
\mathcal{H}_{MCS}=\frac{1}{48\pi M} (\vec{e}^2+b^2) -\frac{1}{2\pi} A_0 b. 
\label{H-MCS}
\end{equation}

The canonical commutation relations imply the gauge invariant algebra
\begin{equation}
[e_i(\vec{x}),e_j(\vec{x}')]= -i (24\pi M)^2 \frac{k}{2\pi} \epsilon_{ij}  \delta(\vec{x}-\vec{x}')\label{21}
\end{equation}
and
\begin{equation}
[b(\vec{x}),e_i(\vec{x}')]=-i (24\pi M)\epsilon_{ij}\partial_j \delta(\vec{x}-\vec{x}').\label{22}
\end{equation}
As a consistency check, we can see that the constraint \eqref{eq:constraintMCS} commutes with $e_i$ and $b$ and consequently with the Hamiltonian.
Together with the Hamiltonian \eqref{H-MCS}, we will see that the above algebra enables a direct comparison with the canonical structure emerging in the continuum limit of the quantum wires system.

Next we consider the SD model in \eqref{nontrivialphases}, with canonical momenta,
\begin{eqnarray}
\tilde{\Pi}_0=0~~~\text{and}~~~\tilde{\Pi}_i =-\dfrac{1}{4\pi k}\epsilon^{ij}\alpha_j, \label{23}
\end{eqnarray}
which are primary constraints. The corresponding canonical Hamiltonian reads
\begin{eqnarray}
	\mathcal{H}_{SD} = \dfrac{1}{2\pi k }\alpha_0 \left(\epsilon_{ij}\partial_{i}\alpha_j\right)-\frac{\tilde{M}}{2}\left(\alpha_0-A_0\right)^2+\frac{\tilde{M}}{2}\left(\vec{\alpha}-\vec{A}\right)^2,\label{24}
\end{eqnarray} 
where we have defined the mass parameter 
\begin{equation}
\tilde{M}_{\tilde\lambda}\equiv \frac{|\tilde{M}^2|}{\tilde{\lambda}}.\label{241}
\end{equation} 
It is convenient to shift the component $\alpha_0$ as $\alpha_0\rightarrow \alpha_0 +A_0$, leading to
\begin{eqnarray}
	\mathcal{H}_{SD} = \dfrac{1}{2\pi k }(\alpha_0+A_0) (\epsilon_{ij}\partial_{i}\alpha_j)-\frac{\tilde{M}_{\tilde\lambda}}{2}(\alpha_0)^2+\frac{\tilde{M}_{\tilde\lambda}}{2}\left(\vec{\alpha}-\vec{A}\right)^2.\label{25}
\end{eqnarray}

After using the algebraic equation of motion $\frac{d}{dt}\tilde{\Pi}_0=-\frac{\delta}{\delta\alpha_0}H_{SD}=0$, the Hamiltonian $\mathcal{H}_{SD}$ is recast as
\begin{equation}
\mathcal{H}_{SD}= \frac{\tilde{M}_{\tilde\lambda}}{2}\left(\vec{\alpha}-\vec{A}\right)^2+\frac{1}{2\tilde{M}_{\tilde\lambda}} \frac{1}{(2\pi k)^2} (\epsilon_{ij}\partial_{i}\alpha_j)^2+\dfrac{1}{2\pi k }A_0 (\epsilon_{ij}\partial_{i}\alpha_j).
\label{H-SD}
\end{equation}
Finally, the canonical commutation relations can be obtained using the Dirac formalism for second class constraints:
\begin{equation}
[\alpha_{i}(\vec{x}),\alpha_j(\vec{x}')]=-i (2\pi k) \epsilon_{ij} \delta(\vec{x}-\vec{x}'),\label{27}
\end{equation}
offering an alternative interpretation for the continuum limit of the quantum wires system. 

The dual formulation of the Hamiltonians $\mathcal{H}_{MCS}$ and $\mathcal{H}_{SD}$ is the backbone of our analysis on the underlying dualities of the fractional quantum Hall effect.


\section{Quantum Wires Description of Laughlin States}
\label{quantumwiressec}

In this section we discuss the quantum wires description of the Laughlin states. After the bosonization of the fermionic theory, we consider the continuum limit, which enables us to identify an emergent gauge structure that can be connected with MCS and SD Hamiltonians.

\subsection{Quantum Wires Formulation}
\label{subsec:qwformulation}

The quantum wires setup starts with a collection of noninteracting one-dimensional wires supporting electrons in the presence of an external magnetic field. The linearized excitations around the Fermi points are gapless, described by the Hamiltonian
\begin{eqnarray}
	{H}_{0} = v_F\sum\limits_y\int dx\,(\psi^{\dagger}_{L,y}i\partial_x\psi_{L,y}-\psi^{\dagger}_{R,y}i\partial_x\psi_{R,y}),
	\label{freeH}
\end{eqnarray}
where $v_F$ is the Fermi velocity and $y$ labels the different wires.   

There are two types of interactions in such a system, intrawire and interwire, and we would like to model the ones able to destabilize the critical theory \eqref{freeH} and drive the system to the Laughlin phase. 

As the different wires in (\ref{freeH}) do not interact, charge is conserved inside each wire. Furthermore, (\ref{freeH})  is invariant under chiral transformations $\psi_{R/L,y}\rightarrow e^{\pm i \alpha}\psi_{R/L,j}$. These symmetries imply conservation of the currents $J_{R/L,y} \equiv \psi^{\dagger}_{R/L,y}\psi_{R/L,y}$, which in turn can be used to introduce current-current (forward) intrawire interactions of type
\begin{eqnarray}
	{H}_{JJ} = \pi \int dx\sum_y\left\{ \lambda_a[(J_{R,y})^2+(J_{L,y})^2]+2 \lambda_bJ_{R,y}J_{L,y}\right\},
	\label{forwardH}
\end{eqnarray}
where $\lambda_a$ and $\lambda_b$ are coupling constants. Under bosonization, the role of $H_{JJ}$ is to renormalize the kinetic parameters.

Interwire interactions describe more general processes in which electric charge is exchanged between different wires. Charge conservation applies to the system as a whole, not inside each wire, effectively realizing a higher dimensional phase. As shown in the pioneering works of \cite{kane2002fractional,teo2014luttinger}, the interactions responsible for driving the system to the Laughlin phase at the filling fraction $\nu=1/m$ are given by  
\begin{eqnarray}
	{H}^{1/m}_{inter}=-g\int dx\sum\limits_{y=1}^N (\psi^{\dagger}_{L,y+1})^{\frac{m+1}{2}}(\psi_{R,y+1})^{\frac{m-1}{2}}(\psi^{\dagger}_{L,y})^{\frac{m-1}{2}}(\psi_{R,y})^{\frac{m+1}{2}} + H.c..
\label{kaneint}	
\end{eqnarray}

If we further submit the system to a probe external field $A$ to measure Hall responses, the minimally coupled action associated with ${H}_0$  is
\begin{eqnarray}
	S_{0}[A] = \int dt dx\sum\limits_y\left\{ i \psi^\dagger_{R,y}\left(D_0+v_FD_1\right)\psi_{R,y}+i\psi^\dagger_{L,y}\left(D_0-v_FD_1\right)\psi_{L,y}\right\},\label{31}
\end{eqnarray}
where  $D_0\equiv \partial_t-iA_{0,y}$ and $D_1\equiv \partial_x -iA_{1,y}$. 

Fermionic operators at coincident points must be treated carefully, so that they are point-split regularized in a gauge invariant way. In the case of the currents $J_{R/L}$, this amounts to define the gauge-invariant operator through the insertion of a Wilson line along $x$-direction
\begin{equation}
J_{R/L,y}\equiv \lim_{\epsilon_{\parallel}\rightarrow 0}\, \psi^{\dagger}_{R/L,y}(x+\epsilon_{\parallel})\,e^{i \int_{x}^{x+\epsilon_{\parallel}} dx A_{1,y}}\psi_{R/L,j}(x)-\langle \psi^{\dagger}_{R/L,y}(x)\psi_{R/L,y}(x)\rangle,\label{32}
\end{equation}
where $\epsilon_{\parallel}$ is a short-distance cutoff along the wires, and the divergent part of the current operator has been subtracted. A similar treatment can be applied to the ${H}^{1/m}_{inter}$ operator \cite{santos2015fractional}, introducing Wilson lines along the discretized $y$-direction between fermion operators that are raised to the same power. For example, 
\begin{equation}
(\psi^{\dagger}_{L,y+1}\psi_{R,y})^{\frac{m+1}{2}}~\rightarrow ~(\psi^{\dagger}_{L,y+1}\, e^{i\int_y^{y+1} dy A_{2,y}}\psi_{R,y})^{\frac{m+1}{2}}.\label{33}
\end{equation}
More generally, the interactions in $H^{1/m}_{inter}$ are implicitly built with point-splitting along the $x$-direction.

\subsection{Bosonization}
\label{subsec:bosonizedformulation}

Having defined all the ingredients of the quantum wires system, we are ready to proceed with bosonization. This is done through the field redefinition
\begin{equation}
	\label{bosonizationmap}
\begin{array}{ccc}
	\psi_{R,y}=\dfrac{\kappa_y}{\sqrt{2\pi \epsilon_{\parallel}}}e^{i\left(\varphi_y+\theta_y\right)},&&\psi_{L,y}=\dfrac{\kappa_y}{\sqrt{2\pi \epsilon_{\parallel}}}e^{i\left(\varphi_y-\theta_y\right)},
\end{array}
\end{equation}
where $\epsilon_{\parallel}$ is the short-distance cutoff previously introduced  and $\kappa_y$ denotes the Klein factors ensuring that fermions from different wires anticommute. 
The only nontrivial commutation rule between the bosonic fields is given by
\begin{eqnarray}
[\theta_y(x),\varphi_{y'}(x')] = i\pi\delta_{yy'}\Theta(x-x'),
\label{originalcommutation}
\end{eqnarray}
where $\Theta(x-x')$ is the Heaviside function.

Using \eqref{bosonizationmap}, the bosonized action associated to the system $H_0+H_{JJ}$ is written as
\begin{eqnarray}
S[A]= \frac{1}{2\pi} \int dt dx\sum\limits_y\left\{-2\partial_x\theta_y(\partial_t\varphi_y-A_{0,y})-v(\partial_x\varphi_y-A_{1,y})^2-u(\partial_x\theta_y)^2\right\},\label{36}
\end{eqnarray}
where $u$ and $v$ are given in terms of the parameters of the forward interactions \eqref{forwardH},
\begin{equation}
	u=v_F+\lambda_a+\lambda_b\quad\text{and}\quad v=v_F+\lambda_a-\lambda_b.
	\label{parameters}
\end{equation}
The corresponding Hamiltonian reads
\begin{equation}
H[A]=\frac{1}{2\pi}\int dx \sum_y \left\lbrace -2\partial_x\theta_yA_{0,y}+v\left(\partial_x\varphi_y-A_{1,y}\right)^2+u\left(\partial_x\theta_y\right)^2 \right\rbrace,\label{38}
\end{equation}
while the interwire Hamiltonian \eqref{kaneint} is recast as
\begin{eqnarray}
H_{inter}^{1/m} &=& -g\int dx\sum\limits_y\kappa_{y+1}\kappa_ye^{i\left(-\Delta_y\varphi_y+\epsilon_{\perp} A_{2,y}+mS\theta_y\right)} + H.c.\nonumber\\
&=&-\int dx\sum\limits_y g_{y,y+1}\sin \left(-\Delta_y\varphi_y+\epsilon_{\perp}  A_{2,y}+mS\theta_y\right).
\label{originallaughlin}
\end{eqnarray}
Here $\Delta_y\varphi_y\equiv\varphi_{y+1}-\varphi_y$, $S\theta_y\equiv\theta_{y+1}+\theta_y$,  and in the last line we have absorbed the Klein factors into the coupling constants $g_{y,y+1}$. The parameter $\epsilon_{\perp} $ denotes the interwire spacing. 

In the absence of an external background, the field configurations that minimize the potential are the constant ones given by $\varphi_y \equiv\varphi_0$ and $ \theta_y\equiv \theta_0= \frac{\pi}{4m}$. Note that the value of $\varphi_0$ remains unspecified since the Hamiltonian involves only derivatives and differences of $\varphi$'s. It is then simpler to set the expectation value $\varphi_0$ to zero via a field shift. In this case, \eqref{originallaughlin} becomes 
\begin{equation}
H_{inter}^{1/m}=-\int dx\sum\limits_y g_{y,y+1} \cos \left(-\Delta_y\varphi_y+\epsilon_{\perp}  A_{2,y}+mS\theta_y\right),
\label{originallaughlin1}
\end{equation}
where now $\varphi_0=\theta_0=0$.

It is more convenient to rewrite the parameters $u$ and $v$ as
\begin{equation}
u= \mathsf{v}/K ~~~\text{and}~~~ v= \mathsf{v} K ,\label{41}
\end{equation}
with
\begin{eqnarray}
	\label{eq:qwparameters}
\mathsf{v}\equiv\sqrt{(v_F+\lambda_a)^2-\lambda_b^2}~~~\text{and}~~~
K\equiv\sqrt{\frac{v_F+\lambda_a-\lambda_b}{v_F+\lambda_a+\lambda_b}}.
\label{newparameters}
\end{eqnarray}
We can then rescale the fields $\varphi$ and $\theta$, 
\begin{equation}
\varphi_y \rightarrow \varphi_y/\sqrt{K}  ~~~\text{and}~~~  \theta_y\rightarrow \sqrt{K}\theta_y,\label{43}
\end{equation}
bringing  the Hamiltonian to the canonical normalization. Note that these rescalings do not change the commutator \eqref{originalcommutation}. Finally, the complete Hamiltonian becomes
\begin{eqnarray}
\label{eq:hamiltonianbosonizedqw}
H&=&\frac{1}{2\pi}\int dx \sum_y \left\{ -2\sqrt{K}\partial_x\theta_yA_{0,y}+\mathsf{v}(\partial_x\varphi_y- \sqrt{K} A_{1,y})^2+\mathsf{v}\left(\partial_x\theta_y\right)^2 \right\}\nonumber \\  
&-&\int dx\sum\limits_y g_{y,y+1} \cos \left(\Delta_y\varphi_y /\sqrt{K}-\epsilon_{\perp} A_{2,y}-m \sqrt{K}S\theta_y\right).
\end{eqnarray}

The strongly coupled limit of the quantum wires system is expected to realize the Laughlin state. A key feature of the bosonized theory is that both strongly coupled and continuum
limits can be simultaneously taken, leading to a simple structure that may be readily compared
with the continuum theories of Sec. \ref{macroscopicsec}. 
In the next subsections we summarize the results of \cite{fontana2019quantum} and \cite{imamura2019coupled}, which arrive
at the description of Laughlin series in terms of the Maxwell-Chern-Simons theory starting
from the quantum wires approach and introducing suitable prescriptions of microscopic degrees
of freedom in terms of emergent gauge fields. The prescriptions in these two papers differ but ultimately connected through particle-vortex duality, as we will clarify soon.


\subsection{Strongly Coupled Continuum Limit I: Maxwell-Chern-Simons Theory}\label{mcsd}

We start with the discussion of \cite{fontana2019quantum}, with the Hamiltonian in \eqref{eq:hamiltonianbosonizedqw} and the following identifications\footnote{Note that the normalization constants used in \cite{fontana2019quantum} are slightly different from those used here.}:
\begin{subequations} \label{eq:bee}
\begin{eqnarray}
b_y &\equiv& \dfrac{2\sqrt{K}}{\epsilon_\perp}\partial_x\theta_y,\\
e_{1,y}&\equiv&2\pi g\left(\frac{1}{\sqrt{K}}\Delta_y\varphi_y-m\sqrt{K}S\theta_y-\epsilon_{\perp}A_{2,y}\right),\\
e_{2,y}&\equiv&-\frac{2\mathsf{v}\sqrt{K}}{\epsilon_{\perp}}\left(\partial_x\varphi_y-\sqrt{K} A_{1,y}\right).
\end{eqnarray}
\end{subequations}

This identification can be tested via their canonical  algebra and Hamiltonian. The Hamiltonian \eqref{eq:hamiltonianbosonizedqw} written in terms of $(e,b)$ becomes
\begin{eqnarray}
\label{80}
H&=&\int dx \sum_j\epsilon_{\perp}\left\{ -\frac{1}{2\pi}b_yA_{0,y}+\frac{1}{8\pi}\left(\frac{1}{\mathsf{v}\Lambda_K}e^2_{2,y}+\frac{\mathsf{v}}{\Lambda_K}b^2_y+\frac{1}{\Lambda_g}e^2_{1,y}\right) + \cdots\right\},
\end{eqnarray}
where we have defined the energy scales 
\begin{equation}
\Lambda_K \equiv K\epsilon_\perp^{-1}~~~\text{and}~~~\Lambda_{g}\equiv \pi \epsilon_\perp g_{y,y+1}.
\end{equation}
The energy scale $\Lambda_K $ can be interpreted as a gap for excitations propagating along the wires, because it involves the forward coupling constants. For similar reasoning, $\Lambda_g$ can be interpreted as a gap for excitations propagating in the perpendicular direction. The ellipsis stand for terms that are either constant or higher-order in the cosine expansion.

The identifications in \eqref{eq:bee} imply the constraint
\begin{equation}
\frac{1}{2\Lambda_g} \partial_x e_{1,y}+\frac{1}{2 \mathsf{v}\Lambda_K} \frac{\Delta_y}{\epsilon_{\perp}} e_{2,y} + \frac12 m Sb_y+B_y=0.
\label{const}
\end{equation}
The $(e,b)$ algebra can be explicitly computed  using \eqref{originalcommutation}, and is given by
\begin{eqnarray}
\left[e_{1,y}\left(x\right),e_{2,y^\prime}\left(x^\prime\right)\right]&=&-i4\pi m\mathsf{v}\Lambda_g\Lambda_K\frac{1}{\epsilon_{\perp}}\left(\delta_{y,y^\prime}+\delta_{y,y^\prime+1}\right)\delta\left(x-x^\prime\right),\\
\left[b_{y}(x),e_{1,y^\prime}\left(x^\prime\right)\right]&=&-i4\pi\Lambda_g\frac{\Delta_y}{\epsilon_{\perp}}\frac{1}{\epsilon_{\perp}}\delta_{y,y^\prime}\delta\left(x-x^\prime\right),\\
\left[b_{y}(x),e_{2,y^\prime}\left(x^\prime\right)\right]&=&i4\pi\mathsf{v}\Lambda_K\frac{1}{\epsilon_{\perp}}\delta_{y,y^\prime}\partial_x\delta\left(x-x^\prime\right).
\label{81}
\end{eqnarray}
To consider the continuum limit we take neighboring wires to be infinitesimally close, i.e. $\epsilon_\perp\rightarrow 0$. This limit is only regular if we also take $K\rightarrow 0$ and $g_{y,y+1}\rightarrow\infty$, setting the respective scales to constant values. From equation \eqref{eq:qwparameters}, we see that the limit $K\to 0$ corresponds to $\lambda_a,\lambda_b\rightarrow \infty$. Therefore, the continuum limit $\epsilon_\perp\rightarrow 0$ is consistent with the strongly coupled regime. Besides, $\sum_y\epsilon_\perp$ is identified as an integral over the perpendicular direction $y$, such that $\frac{\delta_{yy'}}{\epsilon_\perp}\rightarrow\delta(y-y')$.

We can see from the Hamiltonian (\ref{80}) that a spatially isotropic two-dimensional phase emerges when 
\begin{equation}
\Lambda_{g}= \mathsf{v}\Lambda_K. 
\label{iso}
\end{equation}
To compare with the relativistic theories in subsection \ref{subsec:hamiltoniananalysis} we further set $\mathsf{v}=1$. Together with the continuum limit discussed above, the Hamiltonian (\ref{80}) becomes
\begin{eqnarray}
H&=&\int d^2x\left[ -\frac{1}{2\pi}bA_{0}+\frac{1}{8\pi\Lambda_K}\left(\vec{e}^2+b^2\right)-\frac{a_{0}}{2\pi}\left(\frac{1}{2\Lambda_K}\nabla\cdot \vec{e}+m b+B\right)\right],
\label{82}
\end{eqnarray}
where we have used an auxiliary field $a_0$ to impose the constraint (\ref{const}) dinamically. The algebra reduces to
\begin{eqnarray}
\left[e_{i}\left(\vec{r}\right),e_{j}\left(\vec{r}^\prime\right)\right]&=&-i\left(4\pi\Lambda_K\right)^2\frac{m}{2\pi}\epsilon_{ij}\delta\left(\vec{r}-\vec{r}^\prime\right),\label{83}\\
\left[b(\vec{r}),e_{i}\left(\vec{r}^\prime\right)\right]&=&-i4\pi\Lambda_K\epsilon_{ij}\partial_j\delta\left(\vec{r}-\vec{r}^\prime\right).
\label{84}
\end{eqnarray}

Therefore, a direct comparison with the respective objects in subsection \ref{subsec:hamiltoniananalysis} leads to
\begin{equation}
m=k~~~\text{and} ~~~ \Lambda_K=6M.\label{85}
\end{equation}
Using all these identifications, the continuum bosonic quantum wires system explicitly realizes the MCS theory.


\subsection{Strongly Coupled Continuum Limit II: Self-Dual Theory\label{SD}}

The work of \cite{imamura2019coupled} starts with the Hamiltonian \eqref{eq:hamiltonianbosonizedqw} and makes the following field identifications:
\begin{equation}
\alpha_{1,y} \equiv \frac{1}{\sqrt{K}}\partial_x\varphi_y~~~\text{and}~~~\alpha_{2,y}\equiv \frac{1}{\epsilon_{\perp}\sqrt{K}}\left( \Delta_y\varphi_y-m K S\theta_y\right).
\label{sdi}
\end{equation}
Using \eqref{originalcommutation}, it is easy to show they satisfy the algebra
\begin{equation}
[\alpha_{1,y}(x),\alpha_{2,y'}(x')]=-\frac{i \pi m }{\epsilon_{\perp}}(\delta_{y,y'+1}+\delta_{y,y'})\delta(x-x').
\end{equation}

The associated magnetic field $b_{\alpha} \equiv \epsilon_{ij}\partial_i\alpha_{j}$ is cast as
\begin{eqnarray}
b_{\alpha,y} = -2\pi m \rho_y,
\end{eqnarray}
where $\rho_y$ is the two-dimensional electron density,
\begin{eqnarray}
	\rho_y = \dfrac{1}{2\pi\epsilon_\perp}\sqrt{K}\partial_x S\theta_y.
\end{eqnarray}
The expression for $b_\alpha$ in terms of $\rho$ resembles the flux-attachment condition imposed by the Chern-Simons field, with magnetic field determined by the particle density.

Following the same strategy of the previous subsection, we rewrite the Hamiltonian \eqref{eq:hamiltonianbosonizedqw} in terms of $\alpha$,
\begin{equation}
	H = \!\!\!\int\!\!\! dx\sum\limits_y\epsilon_\perp \left\{-A_{0,y}\rho_y +\frac{\mathsf{v}\Lambda_K}{2\pi}[(\alpha_{1,y}-A_{1,y})^2+\frac{\Lambda_{g}}{\mathsf{v}\Lambda_K}(\alpha_{2,y}-A_{2,y})^2] +\frac{\mathsf{v}}{2\pi}\Lambda_K\pi^2 4\rho_y^2\right\},
\end{equation}
and take its strongly coupled continuum limit,
\begin{eqnarray}
	H =\!\!\!\int\!\!\! d^2 x  \left\{-A_0\rho +\frac{\mathsf{v}\Lambda_K}{2\pi}\left[(\alpha_1 -A_1)^2+\frac{\Lambda_g}{\mathsf{v}\Lambda_K}(\alpha_2-A_2)^2\right]	+\frac{\mathsf{v}}{2\pi}\frac{\pi^2}{\Lambda_K}\frac{1}{(2\pi m)^2}(\epsilon_{ij}\partial_i \alpha_j)^2\right\}.\nonumber\\
\end{eqnarray}
The algebra becomes
\begin{equation}
	[\alpha_i(\vec{x}),\alpha_j(\vec{x}')] = -2\pi i m \epsilon_{ij}\delta(\vec{x}-\vec{x}').
\end{equation}
We see that the spatially isotropic case is given precisely by (\ref{iso}), and we also set here $\mathsf{v}=1$. Once more, comparing  these results with their counterpart in the subsection \ref{subsec:hamiltoniananalysis} leads to
\begin{eqnarray}
m=k~~~\text{and}~~~ \Lambda_K = \pi\tilde{M}_{\tilde\lambda}. 
\label{rp}
\end{eqnarray}
Through \eqref{85} we again find the relation between the macroscopic parameters in (\ref{eq:macroscopicrelationofMs}). Therefore, the MCS/SD duality can be viewed as a macroscopic manifestation of the quantum wires description of the Laughlin states of the fractional quantum Hall effect. The duality is embodied by the underlying identification of the microscopic degrees of freedom in the bosonized description with the gauge field or its field-strength.

Finally, let us briefly comment on the relations between the identifications in the MCS and SD models with the quantum wires and the corresponding continuum theories. A direct comparison leads to the following relation between the gauge fields in the two models:
\begin{equation}
e_{1,j}=2 \Lambda_K (\alpha_2-A_2),~~~ e_{2,j}=- 2 \Lambda_K (\alpha_1-A_1),~~~b_a = -\frac{1}{m} b_{\alpha}. 
\label{eq:MCS-SD-id}
\end{equation}
We can immediately see that these equations constitute a solution to the constraint \eqref{eq:constraintMCS} in the MCS theory. This provides an interesting perspective: we can think of the SD model as emerging from the MCS model written in terms of variables that automatically solve the constraint.


\section{From Quantum Wires to Effective Gauge Theories\label{eft}}

In this section, we show that the two prescriptions reviewed in the previous section, relating quantum wires variables to gauge fields, can be naturally obtained from the quantum wires formalism when embedding the model \eqref{eq:hamiltonianbosonizedqw} in a gauge invariant theory.


\subsection{Self-Dual Model}\label{123}

In order to show how the SD model can be obtained as a low-energy limit of the theory \eqref{eq:hamiltonianbosonizedqw} and then justify the prescriptions of section \ref{SD}, we start with its corresponding action:
\begin{eqnarray}
\label{45}
S&=&\int dtdx\sum_y \left\{-\frac{1}{\pi}\partial_x\theta_y\left(\partial_t\varphi_y-\sqrt{K}A_{0,y}\right)-\frac{\mathsf{v}}{2\pi}(\partial_x\varphi_y- \sqrt{K} A_{1,y})^2-\frac{\mathsf{v}}{2\pi}\left(\partial_x\theta_y\right)^2 \right\} \nonumber\\ &+&\int dtdx\sum\limits_y g_{y,y+1} \cos \left(\Delta_y\varphi_y /\sqrt{K}-\epsilon_{\perp} A_{2,y}-m \sqrt{K}S\theta_y\right).
\end{eqnarray}
The charge density couples to the $A_0$ component of the external field and is given by $\rho_y=\frac{\sqrt{K}}{\pi\epsilon_{\perp}}\partial_x\theta_y$.

The action (\ref{45}) can be seen as the low-energy limit of a complex scalar field coupled to an external gauge field $A_\mu$ and a statistical gauge field $\alpha_\mu$ in a specific gauge. The latter is responsible for the flux-attachment mechanism. Indeed, by choosing the gauge $\alpha_{1,y}=0$, and making the identification $\alpha_{2,y}=-\frac{m\sqrt{K}}{\epsilon_{\perp}}S\theta_y$, we obtain the usual relation between magnetic flux and density implied by the flux-attachment:
\begin{eqnarray}
	\label{46.1}
b_y=\partial_x\alpha_{2,y}-\partial_y\alpha_{1,y}=-\frac{m\sqrt{K}}{\epsilon_{\perp}}\partial_xS\theta_y
=-\pi mS\rho_y.
\end{eqnarray}
This identification can be implemented using a Lagrange multiplier $\alpha_0$ in the action,
\begin{eqnarray}
S&=&\int dtdx\sum_y \left\{-\frac{1}{\pi}\partial_x\theta_y\left(\partial_t\varphi_y-\sqrt{K}A_{0,y}\right)-\frac{\mathsf{v}}{2\pi}(\partial_x\varphi_y- \sqrt{K} A_{1,y})^2-\frac{\mathsf{v}}{2\pi}\left(\partial_x\theta_y\right)^2 \right\} \nonumber\\
&+&\int dtdx\sum\limits_y g_{y,y+1} \cos \left(\Delta_y\varphi_y /\sqrt{K}-\epsilon_{\perp} A_{2,y}+\epsilon_{\perp} \alpha_{2,y}\right)
\nonumber\\
&+&\int dtdx\sum\limits_y \frac{\epsilon_{\perp}}{2\pi m}\partial_x \alpha_{0,y}\left(\alpha_{2,y}+\frac{m\sqrt{K}}{\epsilon_{\perp}}S\theta_y\right).
\label{47}
\end{eqnarray}
Gauge invariance can be made explicit after integrating by parts the last line, 
\begin{eqnarray}
	\label{48}
S&=&\int dtdx\sum_y \left\{-\frac{1}{\pi}\partial_x\theta_y\left(\partial_t\varphi_y-\sqrt{K}A_{0,y}+\frac{\sqrt{K}}{2}S\alpha_{0,y}\right)-\frac{\mathsf{v}}{2\pi}(\partial_x\varphi_y- \sqrt{K} A_{1,y}+\sqrt{K} \alpha_{1,y})^2\right\} \nonumber\\ 
&+&\int dtdx\sum\limits_y \left\{-\frac{\mathsf{v}}{2\pi}\left(\partial_x\theta_y\right)^2 +g_{y,y+1} \cos \left(\Delta_y\varphi_y /\sqrt{K}-\epsilon_{\perp} A_{2,y}+\epsilon_{\perp} \alpha_{2,y}\right)\right\}.\nonumber\\
&+&\int dtdx\sum_y \left\{ -\frac{\epsilon_{\perp}}{4\pi m}\epsilon^{\mu\nu\rho}\alpha_{\mu,y}\partial_\nu \alpha_{\rho,y}\right\}.
\end{eqnarray}
Finally, the field $\theta$ can be integrated out, leading to
\begin{eqnarray}
	\label{50}
S&=&\int dtdx\sum_y \left\{\frac{1}{2\mathsf{v}\pi}\left(\partial_t\varphi_y-\sqrt{K}A_{0,y}+\frac{\sqrt{K}}{2}S\alpha_{0,y}\right)^2-\frac{\mathsf{v}}{2\pi}(\partial_x\varphi_y- \sqrt{K} A_{1,y}+\sqrt{K}\alpha_{1,y})^2\right\} \nonumber\\ 
&+&\int dtdx\sum\limits_y \left\{g_{y,y+1} \cos \left(\Delta_y\varphi_y /\sqrt{K}-\epsilon_{\perp} A_{2,y}+\epsilon_{\perp} \alpha_{2,y}\right)-\frac{\epsilon_{\perp}}{4\pi m}\epsilon^{\mu\nu\rho}\alpha_{\mu,y}\partial_\nu\alpha_{\rho,y}\right\}.
\end{eqnarray}
This is the low-energy limit of a Higgs phase of a complex scalar field coupled to a gauge field. To obtain the SD model we simply consider the unitary gauge $\varphi=0$ and then take both the continuum and isotropic limits of this action. Proceeding as in the previous section, the continuum limit reads
\begin{eqnarray}
\label{51}
S&=&\!\!\int\!\! d^3x\left[\frac{\Lambda_K}{2\pi\mathsf{v}}\left(A_{0}-\alpha_{0}\right)^2-\frac{\mathsf{v}\Lambda_K}{2\pi}(A_{1}-\alpha_{1})^2-\frac{\Lambda_g}{2\pi} \left(A_{2}-\alpha_{2}\right)^2-\frac{1}{4\pi m}\epsilon^{\mu\nu\rho}\alpha_{\mu}\partial_\nu \alpha_{\rho}\right].
\end{eqnarray}
After considering the isotropic situation, $\Lambda_K=\Lambda_g$ and $\mathsf{v}=1$, and comparing this action with the right hand side of (\ref{nontrivialphases}), we recover the relations in (\ref{rp}).


\subsection{Maxwell-Chern-Simons: Particle-Vortex Duality\label{MCS}}

We have shown that the effective field theory for the Laughlin series can be naturally embedded in a gauge invariant theory, which coincides with the SD model upon a specific gauge choice. Next we will show that we can achieve the alternative MCS theory using the particle-vortex duality on the wires system in 1+1 dimensions.

It will now be convenient to consider the action \eqref{48} in the gauge $\alpha_2=0$. After integrating out the auxiliary field $\alpha_0$, we are left with
\begin{eqnarray}
	S&=&\int dtdx\sum_y \left\{-\frac{1}{\pi}\partial_x\theta_y\left(\partial_t\varphi_y-\sqrt{K}A_{0,y}\right)-\frac{\mathsf{v}}{2\pi}\left(\partial_x\varphi_y- \sqrt{K}A_{1,y}-mK\partial_x\Delta^{-1}S\theta\right)^2\right\} \nonumber \\
&+&\int dtdx\sum\limits_y \left\{-\frac{\mathsf{v}}{2\pi}\left(\partial_x\theta_y\right)^2 +g_{y,y+1} \cos \left(\Delta_y\varphi_y /\sqrt{K}-\epsilon_{\perp} A_{2,y}\right)\right\}\nonumber\\ 
&+&	\int dtdx\sum_y\left\{ - \frac{\mathsf{u}-\mathsf{v}}{8\pi}\left[\Delta_y\left(\partial_x\varphi_y-\sqrt{K} A_{1,y}\right)\right]^2\right\}.
\label{61}
\end{eqnarray}
We have added an extra term with coupling $\mathsf{u}-\mathsf{v}$ for convenience. It does not change  the qualitative features of the model but makes the quantum wires description more transparent. After some rearrangements, this action can be recast as
\begin{eqnarray}
\label{62}
S&=&\int dtdx\sum_y \left\{-\frac{1}{\pi}\partial_x\theta_{y}\partial_t\varphi_{y}+\frac{\sqrt{K}}{\pi}\partial_x\theta_{y}A_{0,y}-\frac{
		\mathsf{v}}{8\pi}\left(S\left(\partial_x\varphi_y-\sqrt{K}A_{1,y}\right)\right)^2\right\}\nonumber\\ 
&+&\int dtdx\sum\limits_y\left\{-\frac{
		\mathsf{v}m^2K^2}{2\pi}\left(\partial_x\Delta_y^{-1}S\theta\right)^2+\frac{
		\mathsf{v}mK}{\pi}\left(\partial_x\varphi_y-\sqrt{K}A_{1,y}\right)\partial_x\Delta_y^{-1}S\theta-\frac{
		\mathsf{v}}{2\pi}\left(\partial_x\theta_y\right)^2\right\}\nonumber\\
&+&\int dtdx\sum\limits_y \left\{g_{y,y+1} \cos \left(\Delta_y\varphi_y /\sqrt{K}-\epsilon_{\perp} A_{2,y}\right)-\frac{
		\mathsf{u}}{8\pi}\left(\partial_x\Delta_y\varphi_y-\sqrt{K}A_{1,y}\right)^2\right\}.
\end{eqnarray}

Now we turn to the dual theory through the particle-vortex transformation, which is explicit in terms of quantum wires variables. According to \cite{Mross:2015idy, mross2017symmetry}, the particle-vortex transformation has the form
\begin{eqnarray}
	\tilde{\varphi}_{y-1/2}&=&\sum\limits_{y'}\text{sign}(y'-y+1/2)\theta_{y'}=-2\Delta_y^{-1}\theta_{y},\label{621}\\
	\tilde{\theta}_{y+1/2} &=&\dfrac12\Delta_y\varphi_{y}\label{622}.
\end{eqnarray}
The second equality in (\ref{621}) follows form the identity $\sum_{y'}\left(\Delta_y\right)_{yy'}\text{sign}(y'-y''+1/2)=2\delta_{yy''}$, with $\left(\Delta_y\right)_{yy'}=\delta_{y+1,y'}-\delta_{y,y'}$ being the matrix elements of the $\Delta_y$ operator. The inverse transformations are given by
\begin{eqnarray}
	\varphi_{y} &=& 2\Delta_y^{-1}\tilde{\theta}_{y+1/2},\label{624}\\
	\theta_{y} &=&-\frac{1}{2}\Delta_y\tilde{\varphi}_{y-1/2}\label{625}.
\end{eqnarray}
 Applying these transformations to the action (\ref{62}), we obtain  
 \begin{eqnarray}
 	S&=&\int dtdx\sum_y \left\{-\frac{1}{\pi}\partial_x\tilde{\theta}_{y+1/2}\partial_t\tilde{\varphi}_{y+1/2}+\frac{\sqrt{K}}{2\pi}\partial_x\Delta_y^T\tilde{\varphi}_{y+1/2}A_{0,y}-\frac{\mathsf{v}m^2K^2}{8\pi}\left(\partial_xS^T\tilde{\varphi}_{y+1/2}\right)^2\right\}\nonumber\\\nonumber
 	&+&\int dtdx\sum\limits_y\left\{-\frac{\mathsf{v}mK}{\pi}\left(\partial_x\Delta_y^{-1}\tilde{\theta}_{y+1/2}-\frac{1}{2}\sqrt{K}A_{1,y}\right)\left(\partial_xS^T\tilde{\varphi}_{y+1/2}\right)\right\}\\ \nonumber 
 	&+&\int dtdx\sum\limits_y\left\{-\frac{\mathsf{v}K}{2\pi}\left(S\left(\partial_x\tilde{\theta}_{y+1/2}-\frac{1}{2}\sqrt{K}\Delta_y A_{1,y}\right)\right)^2-\frac{\mathsf{u}}{2\pi}\left(\partial_x\tilde{\theta}_{y+1/2}-\frac{1}{2}\sqrt{K}\Delta_y A_{1,y}\right)^2\right\}\\ 
 	&+&\int dtdx\sum\limits_y \left\{g_{y,y+1} \cos \left(2\tilde{\theta}_{y+1/2}/\sqrt{K}-\epsilon_{\perp} A_{2,y}\right)-\frac{\mathsf{v}}{8\pi}\left(\partial_x\Delta_y^T\tilde{\varphi}_{y+1/2}\right)^2\right\},
 \label{63}
\end{eqnarray}
which is nonlocal because of the presence of $\Delta_y^{-1}$. However, we show in the Appendix \ref{AA} that this nonlocal action is equivalent to the following local model:
\begin{eqnarray}
\label{64}
S&=&\int dtdx\sum_y \left\{-\frac{1}{\pi}\partial_x\tilde{\theta}_{y+1/2}\partial_t\tilde{\varphi}_{y+1/2}+\frac{\sqrt{K}}{2\pi}\partial_x\Delta_y^T\tilde{\varphi}_{y+1/2}A_{0,y}-\frac{\gamma}{2\pi}\left(\partial_x\tilde{\varphi}_{y+1/2}-\sqrt{K}a_{1,y+1/2}\right)^2\right\}\nonumber\\ 
&+&\int dtdx\sum\limits_y \left\{-\frac{\lambda}{8\pi}\left(\partial_x\Delta_y^T\tilde{\varphi}_{y+1/2}\right)^2-\frac{\omega}{2\pi}\left(\partial_x\tilde{\theta}_{y+1/2}-\frac{\sqrt{K}}{2}\Delta_y A_{1,y}\right)^2\right\} \nonumber \\
&+&\int dtdx\sum\limits_y \left\{g_{y,y+1} \cos \left(2\tilde{\theta}_{y+1/2}/\sqrt{K}-\epsilon_{\perp} A_{2,y}\right)+\frac{m}{4\pi}\Delta_y a_{0,y}Sa_{1,y+1/2}\right\}\nonumber\\
&+&\int dtdx\sum\limits_y \left\{+\frac{1}{\pi K^{3/2}}a_{0,y}\left(\partial_x\tilde{\theta}_{y+1/2}-\frac{\sqrt{K}}{2}\Delta_y A_{1,y}\right)\right\}\nonumber\\
&+&\int dtdx \sum\limits_y \left\{ \frac{1}{8\pi}\left[\frac{\alpha}{\gamma}\left(\Delta_y a_{0,y}\right)^2+\beta \gamma\left(\Delta_y a_{1,y+1/2}\right)^2\right]\right\},
\end{eqnarray}
where the new parameters $\gamma, \lambda, \omega, \alpha, \beta$ are such that
\begin{eqnarray}
\frac{\gamma}{K^2\left(\alpha K+m^2\right)}=\mathsf{v},~~~ \omega+\frac{\gamma}{\alpha K^3}=\mathsf{u}, ~~~\text{and}~~~
\lambda=\mathsf{v}\left(1+k^2m^2\right)\label{741},
\end{eqnarray}
leading to the expected match with \eqref{63} after integration over the emergent gauge fields. The above action describes scalar fields on the wires, minimally coupled to an emergent gauge field in the presence of Chern-Simons and Maxwell terms. We notice that this description is in the $a_2=0$ gauge and there is no $x$-component of the electric field. However, this component will emerge when we integrate over the matter fields.

In order to show that this local action reduces to  Maxwell-Chern-Simons at low energies, we use the first relation in \eqref{741} to recast the local action \eqref{64} in terms of $\gamma$ instead of $\mathsf{v}$:
\begin{eqnarray}
\label{75}
S&=&\int dtdx\sum_y \left\{-\frac{1}{\pi}\partial_x\tilde{\theta}_{y+1/2}\partial_t\tilde{\varphi}_{y+1/2}+\frac{\sqrt{K}}{2\pi}\partial_x\Delta_y^T\tilde{\varphi}_{y+1/2}A_{0,y}-\frac{\gamma}{2\pi}\left(\partial_x\tilde{\varphi}_{y+1/2}-\sqrt{K}a_{1,y+1/2}\right)^2\right\} \nonumber\\ 
&+&\int dtdx\sum\limits_y \left\{-\frac{\gamma}{8\pi K^2}\frac{1-K^2m^2}{\alpha K+m^2}\left(\partial_x\Delta_y^T\tilde{\varphi}_{y+1/2}\right)^2-\frac{\mathsf{u}\alpha K^3-\gamma}{2\pi \alpha K^3}\left(\partial_x\tilde{\theta}_{y+1/2}-\frac{\sqrt{K}}{2}\Delta_y A_{1,y}\right)^2\right\} \nonumber\\
&+&\int dtdx\sum\limits_y \left\{g_{y,y+1} \cos \left(2\tilde{\theta}_{y+1/2}/\sqrt{K}-\epsilon_{\perp} A_{2,y}\right)+\frac{m}{4\pi}\Delta_y a_{0,y}Sa_{1,y+1/2}\right\} \nonumber\\
&+&\int dtdx\sum\limits_y \left\{\frac{1}{\pi K^{3/2}}a_{0,y}\left(\partial_x\tilde{\theta}_{y+1/2}-\frac{\sqrt{K}}{2}\Delta_y A_{1,y}\right)\right\}\nonumber\\
&+&\int dtdx\sum\limits_y \left\{\frac{1}{8\pi}\left[\frac{\alpha}{\gamma}\left(\Delta_y a_{0,y}\right)^2+\beta \gamma\left(\Delta_y a_{1,y+1/2}\right)^2\right]\right\}.
\end{eqnarray}

Towards the continuum limit, we expand the cosine interaction up to quadratic terms and integrate out the $\theta$. We disregard terms with more than two derivatives since they are irrelevant at low energies. In addition, we rescale the fields $a_0$ and $a_1$ to $Ka_0$ and $\frac{1}{K}a_1$, without affecting the Chern-Simons coefficient, and reintroduce the $a_2$ component to restore the gauge invariance of the model. We then have 
\begin{eqnarray}
\label{76}
S&=&\int dtdx\sum\limits_y\left\{-\frac{\gamma}{2\pi}\left(\partial_x\tilde{\varphi}_{y+1/2}-\frac{1}{\sqrt{K}}a_{1,y+1/2}\right)^2+\frac{\alpha K^2\left(\epsilon_{\perp}\right)^2}{8\pi \gamma}\left(e_{2,y}\right)^2+\frac{\beta \gamma\left(\epsilon_{\perp}\right)^2}{8\pi K^2}\left(b_{y+1/2}\right)^2\right\}\nonumber\\ \nonumber
&+&\int  dtdx\sum\limits_y \left\{\frac{K}{8g\pi}\left(\partial_t\partial_x\tilde{\varphi}_{y+1/2}-\frac{1}{\sqrt{K}}\partial_xa_0\right)^2+\frac{\epsilon_{\perp}}{2\pi}\left(\partial_t\partial_x\tilde{\varphi}_{y+1/2}-\frac{1}{\sqrt{K}}\partial_xa_{0,y}\right)\sqrt{K}A_{2,y}\right\}\\ \nonumber
&+& \int dtdx\sum\limits_y \left\{\frac{m\epsilon_{\perp}}{8\pi}\epsilon^{\mu\nu\rho}\left(Sa_{\mu}\right)\partial_\nu a_{\rho}+\frac{\epsilon_{\perp}}{2\pi}A_{1,y}\left(\frac{1}{\epsilon_{\perp}}\Delta_y a_{0,y-1}-\partial_ta_{2,y}\right)\right\}\\ 
&+&\int dtdx\sum\limits_y \left\{\frac{\sqrt{K}}{2\pi}\left(\Delta_y\tilde{\varphi}_{y-1/2}-\frac{1}{\sqrt{K}}a_{2,y-1/2}\right)\partial_xA_{0,y}\right\},
\end{eqnarray}
where $e_{2,y}=\partial_ta_2-\partial_ya_0$, $b_{y+1/2}=\partial_xa_2-\partial_ya_1$, and $\partial_y\equiv\frac{\Delta_y}{\epsilon_\perp}$. 
Integrating out the $\tilde{\varphi}$-field and retaining only two derivatives terms, we finally obtain
\begin{eqnarray}
\label{77}
S&=&\int dtdx\sum\limits_y\left\{\frac{\alpha K^2\left(\epsilon_{\perp}\right)^2}{8\pi \gamma}\left(e_{2,y}\right)^2+\frac{\beta \gamma\left(\epsilon_{\perp}\right)^2}{8\pi K^2}\left(b_{y+1/2}\right)^2+\frac{1}{8g\pi ^2}\left(e_{1,y}\right)^2\right\}\\ \nonumber
&+&\int dtdx\sum\limits_y \left\{+\frac{m\epsilon_{\perp}}{8\pi}\epsilon^{\mu\nu\rho}\left(Sa_{\mu}\right)\partial_\nu a_{\rho}+\frac{\epsilon_{\perp}}{2\pi}\epsilon^{\mu\nu\rho}A_{\mu,y}\partial_\nu a_{\rho}\right\}.
\end{eqnarray}
This is an anisotropic version of the MCS model on the wires system. For the continuum limit, we consider as before $\epsilon_{\perp}\rightarrow 0$, $K\rightarrow 0$, and $g\rightarrow\infty$, while keeping $K\epsilon^{-1}_\perp\equiv\Lambda_K$ and $\pi g\epsilon_\perp\equiv\Lambda_g$ fixed. To have finite Maxwell terms, we also consider that $\alpha$ and $\beta$ go to infinity as $1/K$, i.e., given constants $c$ and $d$, we take $\alpha=c/K$ and $\beta=-d/K$. With these choices, we obtain 
\begin{eqnarray}
\label{78}
S&=&\int d^3x\left\{\frac{1}{8\pi \tilde{\mathsf{v}}\Lambda_K}\left(e_{2}\right)^2-\frac{cd\tilde{\mathsf{v}}}{8\pi\Lambda_K}b^2+\frac{1}{8\pi\Lambda_g}\left(e_{1}\right)^2\right\}\\ \nonumber
&+&\int d^3x \left\{\frac{m}{4\pi}\epsilon^{\mu\nu\rho}a_\mu\partial_\nu a_\rho+\frac{1}{2\pi}\epsilon^{\mu\nu\rho}A_{\mu}\partial_\nu a_{\rho}\right\},
\end{eqnarray}
where the renormalized velocity is $\tilde{\mathsf{v}}\equiv \frac{\mathsf{v}\left(c+m^2\right)}{c}$. Now, in order to compare it with the MCS theory in \ref{subsec:hamiltoniananalysis}, we  consider again the isotropic limit of this action, which is attained by making $\Lambda_K=\Lambda_{g}=6M$, $d=1/c$ and $\tilde{\mathsf{v}}=1$.


\subsection{A More Direct Route to Maxwell-Chern-Simons Theory\label{mw}}

As discussed in section \ref{mcsd}, the same effective theory \eqref{78} was obtained in \cite{fontana2019quantum} using quantum wires, but relating the scalar field $\varphi$ with components of the field strength of the emergent gauge fields. This can be seen as a shortcut of the above description, leading to the microscopic theory directly to the low-energy effective model. Here instead, we have attained the low-energy theory \eqref{78} using the particle-vortex transformation in the wires language and only after taking the low-energy limit. Therefore, it would be interesting to recast the results of \cite{fontana2019quantum} in a more convenient language, making explicit the connection with particle-vortex duality. To this end, we first obtain the relation between the canonical momenta and gauge fields from \eqref{82}-\eqref{84},
\begin{eqnarray}
\Pi_{1,y}&=&\frac{1}{4\pi\Lambda_g}e_{1,y}+\frac{1}{2\pi}A_{2,y}+\frac{m}{4\pi}a_{2,y}\label{86}\\
\Pi_{2,y}&=&\frac{1}{4\pi\mathsf{v}\Lambda_K} e_{2,y}-\frac{1}{2\pi}A_{1,y}-\frac{m}{4\pi}a_{1,y},
\label{87}
\end{eqnarray}
and rewrite the relations \eqref{eq:bee} accordingly, obtaining
\begin{eqnarray}
\Pi_{1,y}+\frac{m}{4\pi}a_{2,y}&=&\frac{1}{2\pi\sqrt{K}\epsilon_{\perp}}\Delta_y\varphi_y\label{88.1}\\
\Pi_{2,y}-\frac{m}{4\pi}a_{1,y}&=&-\frac{1}{2\pi\sqrt{K}}\partial_x\varphi_y\label{88.2}\\
\partial_xa_{2,y}-\frac{1}{\epsilon_{\perp}}\Delta_y a_{1,y}&=&\frac{2\sqrt{K}}{\epsilon_{\perp}}\partial_x\theta_y.
\label{88.3}
\end{eqnarray}
In the gauge $a_{1,y}=0$ we immediately have $a_{2,y}=\tfrac{2\sqrt{K}}{\epsilon_{\perp}}\theta_y$, yielding the constraint
\begin{equation}
\partial_x\Pi_{1,y}+\frac{1}{\epsilon_{\perp}}\Delta_y\Pi_{2,y}+\frac{m}{4\pi}b=0.\label{89}
\end{equation}

The relations above can be seen as a generalization of winding-momentum duality between Maxwell theory and the XY model in 2+1 dimensions, which is an equivalent way to view the particle-vortex duality. Alternatively, we can view  \eqref{88.1}-\eqref{88.3}  as a direct connection between the action \eqref{50}, which is the SD model after choosing the unitary gauge, and the MCS model. This is made explicit after rewriting \eqref{88.1}-\eqref{88.3} in a covariant form,
\begin{eqnarray}
\frac{1}{4\pi\Lambda_K}f_{\mu\nu}-\frac{m}{2\pi}\epsilon_{\mu\nu\rho}a^{\rho}-\frac{1}{2\pi}\epsilon_{\mu\nu\rho}A^{\rho}=-\frac{1}{2\pi\sqrt{\Lambda_K}}\epsilon_{\mu\nu\rho}\partial^\rho\varphi\label{90},
\end{eqnarray}
where we already considered the isotropic and continuum limits ($\Lambda_K=\Lambda_g$ and $\varphi(x,y,t)=\frac{1}{\epsilon_\perp}\varphi_y(x,t)$). To obtain the last relation (\ref{88.3}) from (\ref{90}), we use the equation of motion for $\theta$ in \eqref{61}, given by
\begin{equation}
\partial_x\theta_y=-\frac{1}{\mathsf{v}}\left(\partial_t\varphi_y + \frac{\sqrt{K}}{2}S\alpha_{0,y}-\sqrt{K}A_{0,y}\right),
\end{equation}
and make the identification $a_\mu=-m^{-1}\alpha_\mu$ between the gauge fields in the two models. Furthermore, we can express the inverse of the relation (\ref{90}) as 
\begin{eqnarray}
\frac{1}{4\pi\sqrt{\Lambda_K}}\epsilon^{\mu\nu\rho}f_{\nu\rho}=-\frac{1}{\pi}\left(\partial^\mu\varphi+\sqrt{\Lambda_K}\alpha^{\mu}-\sqrt{\Lambda_K}A^{\mu}\right).\label{91}
\end{eqnarray}

Notice we recover the equations in \eqref{eq:MCS-SD-id}  in the gauge $\varphi=0$. The relations \eqref{90} and \eqref{91} clearly display a winding-momentum structure, which is typical of particle-vortex duality: the equation of motion of the MCS model is obtained taking the divergence of \eqref{90}, which from the form of the left hand side of this equation is identically satisfied in terms of the $\varphi$ variable. Analogously, the equation of motion for $\varphi$ is obtained by taking the divergence of \eqref{91}, which is again trivially satisfied in terms of the gauge field $a_\mu$ from the Bianchi identity.

We conclude this subsection by pointing out that the relation between the prescriptions used in \cite{imamura2019coupled} and \cite{fontana2019quantum} are connected by particle-vortex duality. This can be verified either from the quantum wires approach using the particle-vortex transformation proposed in \cite{mross2017symmetry}, which lead us from (\ref{61}) to (\ref{78}), or via the expressions relating momentum and winding number \eqref{88.1}-\eqref{88.3} or \eqref{90} and \eqref{91}, which directly lead to the low-energy effective theory.


\subsection{Vortex Creation Operators\label{vo}}

In the previous section we have shown that the microscopic Hamiltonian for the Laughlin series of QH states can be macroscopically described  either by the SD or MCS models, and that these descriptions are connected by the particle-vortex duality transformations \eqref{90}-\eqref{91}. This duality was derived both from the microscopic wires in  1+1 dimensions and directly in terms of emergent gauge fields in 2+1 dimensions. Here we furhter explore this view and investigate the vortex creation operators in terms of both wires and emergent gauge fields in the continuum. 

Applying $\Delta_y\partial_x$ to particle vortex relations (\ref{621}) and $\partial_x$ to (\ref{622}), we get
\begin{eqnarray}
	\frac{1}{2\pi}\partial_x\Delta_y\tilde{\varphi}_{y-1/2}&=&p_y\label{96}\\
	\tilde{p}_{y+1/2}&=&-\frac{1}{2\pi}\Delta_y\partial_x\varphi_y,
\label{97}
\end{eqnarray}
where $p_y\equiv-\frac{1}{\pi}\partial_x\theta_y$ and $\tilde{p}_{y+1/2}\equiv-\frac{1}{\pi}\partial_x\tilde{\theta}_{y+1/2}$
are the canonical momentum of $\varphi_y$ and $\tilde{\varphi}_{y+1/2}$, respectively. We will show that $\Delta_y\partial_x\varphi_y$ measures the winding number of a specific vortex configuration of $\varphi$ in 2+1 dimensions with the branch cut of the compact variable $\varphi$ running along the $y$ direction. Thus, (\ref{96}) and $(\ref{97})$ express the momentum-winding duality in the quantum wires variables.

To understand how the winding number can be described this way, we note that the operator 
\begin{equation}
e^{i \tilde{\varphi}_{y+1/2}}=\ldots e^{-i \theta_{y-1}} e^{-i \theta_{y}}e^{i \theta_{y+1}}\ldots
\label{vf1}
\end{equation}
creates a half-phase slip of $-\pi$ in all the wires with $y' \leq y$ and a half-phase slip of $+\pi$ in all the wires with $y'>y$, both for $x^\prime>x$. Since, $\varphi$ is a compact variable, a nontrivial winding configuration around $x,y+1/2$ contains a branch cut starting at this point and extending to infinity along some line through which $\varphi$ is discontinuous. 

For the vortex configuration created by $e^{i\tilde{\varphi}}$, which we will denote by $\lambda_{y^\prime}(x^\prime)$, the branch cut is a straight line along the dual wire $y+1/2$. The circulation of $\lambda_{y^\prime}(x^\prime)$ along a small square centered at the point $x, y+1/2$ in the quantum wires system can be calculated by $\sumcirclearrowleft \eta^i\bar{\Delta}_i\lambda_{y}(x)= \sumcirclearrowleft\bar{\Delta}_x\lambda_y(x)+\bar{\Delta}_y\lambda_y(x)$, with $\eta^i$ being an unitary vector tangent to the closed path and $\bar{\Delta}_i\equiv\Delta_i~\mod 2\pi$. For smooth $\varphi$ configurations, $\bar{\Delta}_i\varphi\equiv \Delta_i\varphi$. Considering the branch cut position for the vortex field $\lambda$, we have $\bar{\Delta}_x\lambda_{y}=\Delta_x\lambda_{y}$ and $\bar{\Delta}_{y^\prime}\lambda_{y^\prime}=\Delta_{y^\prime}\lambda_{y^\prime}-2\pi\Theta(x^\prime-x)\delta_{y,y^\prime}$. Therefore, $\sumcirclearrowleft \eta^i\bar{\Delta}_i\lambda_{y}(x)=\epsilon_{ij}\Delta_i\bar{\Delta}_j\lambda_{y}(x)=\Delta_y\Delta_x\lambda_{y}(x)$. Considering the continuum limit of the relations (\ref{96}) and (\ref{97}) and taking into account our discussion of the mod $2\pi$ derivative, we can rewrite them in a more convenient form as
\begin{eqnarray}
	\frac{1}{2\pi}\epsilon_{ij}\partial_i\bar{\partial}_j\tilde{\varphi}(x,y)&=&p(x,y)\label{99}\\
	\tilde{p}(x,y)&=&-\frac{1}{2\pi}\epsilon_{ij}\partial_i\bar{\partial}_j\varphi(x,y).
\label{100}
\end{eqnarray}
Here $p(x,y)=-\frac{1}{\pi}\partial_x\theta(x,y)$, $\tilde{p}=-\frac{1}{\pi}\partial_x\tilde{\theta}(x,y)$, with $\tilde{\varphi}(x,y)=\lim_{\epsilon_{\perp}\rightarrow 0}\epsilon_{\perp}\tilde{\varphi}_{y+1/2}(x)$ and $\tilde{\theta}(x,y)=\lim_{\epsilon_{\perp}\rightarrow 0}\frac{1}{\epsilon^{3/2}_{\perp}}\tilde{\theta}_{y+1/2}(x)$.

Defining $\Pi_i$ and $a_i$ via
\begin{eqnarray}
 \tilde{p}&=&-\sqrt{\Lambda_K}\left(\nabla\cdot\Pi+\frac{m}{4\pi}\epsilon_{ij}\partial_ia_j\right),\label{101}\\
a_i&=&-\sqrt{\Lambda_K}\bar{\partial}_i\tilde{\varphi},
\end{eqnarray}
 we finally obtain
\begin{eqnarray}
	-\frac{1}{2\pi}\epsilon_{ij}\partial_i a_j&=&\sqrt{\Lambda_K}p(x,y),\label{103}\\
	\Pi_i+\frac{m}{4\pi}\epsilon_{ij}a_j&=&\frac{1}{2\pi \sqrt{\Lambda_K}}\epsilon_{ij}\bar{\partial}_j\varphi(x,y),
\label{104}
\end{eqnarray}
which are the relations \eqref{88.1}-\eqref{88.3} for the winding-momentum duality.

From this discussion and the rescaling (\ref{43}) we see that $e^{i\sqrt{K}\tilde{\varphi}_{y+1/2}(x)}$ creates a vortex of $\varphi_y$ at the dual wire $y+1/2$ and $x^\prime=x$, whereas $e^{\frac{1}{\sqrt{K}}\varphi_y(x)}$ creates a vortex of $\tilde{\varphi}_{y+1/2}$ at the wire $y$ and $x^\prime=x$. We have shown that when taking the continuum limit, the original quantum wires system (\ref{eq:hamiltonianbosonizedqw}) can be identified directly as the self dual model (\ref{51}), and with the MCS model (\ref{78}) using the particle-vortex duality (\ref{621})-(\ref{622}). Therefore, we should be able to show that when written in terms of the emergent gauge fields $\alpha_\mu$ and $a_\mu$, the operators $e^{i\sqrt{K}\tilde{\varphi}_{y+1/2}(x)}$ and $e^{i\frac{1}{\sqrt{K}}\varphi_y(x)}$ create vortices in the SD and MCS models, respectively. 

We start with $e^{i\sqrt{K}\tilde{\varphi}_{y+1/2}(x)}$. From (\ref{96}),
\begin{equation}
	\sqrt{K}\tilde{\varphi}_{y+1/2}(x)=\pi\sqrt{\Lambda_K}\int^x dx^\prime\left(\int^y_{-\infty} dy^\prime p(x^\prime,y^\prime)-\int_y^{\infty} dy^\prime p(x^\prime,y^\prime)\right),\label{105}
\end{equation}
where we took the continuum limit on the right hand side.
Since $p(x,y)=-\frac{1}{\pi}\partial_x\theta(x,y)$ and using (\ref{46.1}), we get
\begin{equation}
	\sqrt{K}\tilde{\varphi}_{y+1/2}(x)=\frac{1}{2m}\int dx^\prime dy^\prime \Theta(x-x^\prime)\left(\Theta(y-y^\prime)-\Theta(y^\prime-y)\right) b_\alpha(x^\prime,y^\prime).\label{106}
\end{equation}
Defining $\lambda(\vec{r}-\vec{r}^\prime)\equiv \pi\Theta(x-x^\prime)[\Theta(y-y^\prime)-\Theta(y^\prime-y)]$, we can write the vortex creating operator in the SD model as
\begin{equation}
	V_{\alpha}(\vec{r})=\exp {\frac{i}{2\pi m}\int d^2\vec{r}^\prime\lambda(\vec{r}-\vec{r}^\prime)b_\alpha(\vec{r}^\prime)}.\label{107}
\end{equation}
From the algebra (\ref{27}) with $k=m$, we can show how the gauge field $\alpha_i$ transforms under the action of $V_\alpha$:
\begin{equation}
	V_{\alpha}(\vec{r})\alpha_i(\vec{r}')V^\dagger_{\alpha}(\vec{r})=\alpha_i(\vec{r}')+\bar{\partial}_i\lambda(\vec{r}-\vec{r}').\label{108}
\end{equation}
The function $\lambda$ has a $2\pi$ discontinuity at $y^\prime=y$ for $x^\prime<x$ and discontinuities $\pi$ or $-\pi$ at $x^\prime=x$ depending if $y>y^\prime$ or $y<y^\prime$. Since the derivative $\bar{\partial}$ is mod $2\pi$, the operator $V_\alpha$ generates singular gauge transformations over $\alpha_i$, such that $\alpha^\prime_i=V_{\alpha}\alpha_iV^\dagger_\alpha$ satisfies
\begin{equation}
	\epsilon_{ij}\partial_i\alpha^\prime_j(\vec{r}')=\epsilon_{ij}\partial_i\alpha_j(\vec{r}')+2\pi\delta(\vec{r}-\vec{r}'),\label{28}
\end{equation}
which corresponds to a creation of a unit of flux at $\vec{r}=\vec{r}'$.

Similarly, we can investigate the vortex creating operator $V_a(\vec{r})=e^{i\frac{1}{\sqrt{K}}\varphi(\vec{r})}$ in the MCS model. From (\ref{100}) and (\ref{101}), we obtain
\begin{equation}
	V_{a}(\vec{r})=\exp {i\int d^2\vec{r}^\prime\lambda(\vec{r}-\vec{r}^\prime)\left(\nabla\cdot\Pi(\vec{r}^\prime)+\frac{m}{4\pi}b_a(\vec{r}^\prime)\right)}.\label{29}
\end{equation}
Then, using the canonical algebra between $\pi_i$ and $a_j$, we obtain
\begin{equation}
	a'_i(\vec{r}')=a_i(\vec{r}')-\bar{\partial}_i\lambda(\vec{r}-\vec{r}'),\label{30}
\end{equation}
which again represents a creation of a negative unit of flux at $\vec{r}'=\vec{r}$.

The above discussion provides an alternative way to understand the identification of the low-energy effective theory for the Laughlin series of QH effect either as the SD or MCS models. In the previous sections we have shown that the SD model can be identified directly in terms of a statistical gauge field, whereas MCS is the effective theory describing low-energy  vortices of the original quantum wires model. And now we have presented the low-energy limit of the particle and vortex creating operators in the quantum wires model and shown that they indeed coincide with the corresponding operators in the effective theories in the low-energy limit.




\section{Final Remarks}\label{discussionsec}

The emergence of gauge fields (hydrodynamical or statistical) is one of the imprints of low-energy effective descriptions of the fractional quantum Hall system, encoding the highly complex internal structure due to the collective behavior of the electrons of the quantum fluid. At first sight it seems to be inconceivable to establish any kind of direct connection between emergent gauge degrees of freedom and the electrons. However, the quantum wires formulation represents a remarkable achievement in this direction, enabling an explicit identification between the original microscopic degrees of freedom and the emergent gauge theory in the infrared limit. In fact, being a microscopical description, the quantum wires are expected to somehow encompass both hydrodynamical and statistical gauge structures.

We have discussed how these structures emerge when simultaneously taking the strong coupling and continuum limits. There are two equivalent identifications between the quantum wires variables and the  gauge fields in the continuum. These two facets become physically transparent when we further consider the vortex dual of the original model and embed both original and dual models in gauge invariant theories. The Maxwell-Chern-Simons and the Self-Dual theories emerge in the IR limit of Wen-Zee and ZHK theories, respectively. Therefore, we have provided a unified framework for describing the Laughlin states of the quantum Hall system. As a by-product of our analysis, we have discussed that the well-known connection between SD and MCS models is essentially a particle-vortex relation.

Given the interesting interplay between quantum wires and the theories in the continuum, leading to a series of relations between different approaches, one is tempted to extend this construction to other quantum Hall phases. In particular, our results might be applied to certain non-Abelian phases, like the Moore-Read phase. Their quantum wires formulation has been discussed in  \cite{teo2014luttinger}, and the corresponding non-Abelian Chern-Simons theories were investigated in \cite{Fradkin:1997ge,Fradkin:1998ii}. We believe it would be enlightening to find explicit relations between such approaches using the type of connections we have discussed here.


\section{Acknowledgments}\label{acknowsec}

This work is partially funded by CNPq.

\appendix

\section{Emergence of Maxwell and Chern-Simons Terms in the Quantum Wires System}\label{AA}

In this Appendix we show how the non-local quantum wires model \eqref{63} can be obtained from the local action \eqref{64} by integrating out the emergent gauge fields $a_0$ and $a_1$ in \eqref{64}. After integrating out $a_0$, we get
\begin{eqnarray}
\label{66}
S&=&\int dtdx\sum_y \left\{-\frac{1}{\pi}\partial_x\tilde{\theta}_{y+1/2}\partial_t\tilde{\varphi}_{y+1/2}+\frac{\sqrt{K}}{2\pi}\partial_x\Delta_y^T\tilde{\varphi}_{y+1/2}A_{0,y}-\frac{\gamma}{2\pi}\left(\partial_x\tilde{\varphi}_{y+1/2}\right)^2\right\}\nonumber\\ \nonumber 
&+&\int dtdx\sum\limits_y\left\{-\frac{\omega}{2\pi}\left(\partial_x\tilde{\theta}_{y+1/2}-\frac{\sqrt{K}}{2}\Delta_y A_{1,y}\right)^2-\frac{2\gamma}{\pi\alpha K^{3}}\left(\Delta_y^{-1T}\left(\partial_x\tilde{\theta}_{y+1/2}-\frac{\sqrt{K}}{2}\Delta_y A_{1,y}\right)\right)^2\right\}\\ \nonumber
&+&\int dtdx\sum\limits_y \left\{-\frac{\lambda}{8\pi}\left(\partial_x\Delta_y^T\tilde{\varphi}_{y+1/2}\right)^2 +g_{y,y+1} \cos \left(2\tilde{\theta}_{y+1/2}/\sqrt{K}-\epsilon_{\perp} A_{2,y}\right)\right\}\\ \nonumber
&+&\int dtdx\sum\limits_y\left\{\frac{\gamma}{\pi}\left(\partial_x\tilde{\varphi}_{y+1/2}-\frac{m}{\alpha K^{2}}S^T\Delta_y^{-1T}\left(\partial_x\tilde{\theta}_{y+1/2}-\frac{\sqrt{K}}{2}\Delta_y A_{1,y}\right)\right)\sqrt{K}a_{1,y+1/2}\right\}\\ 
&+&\int dtdx\sum\limits_y \left\{-\frac{\gamma}{2\pi}\sqrt{K}a_{1,y+1/2}M\sqrt{K}a_{1,y+1/2}\right\},
\end{eqnarray}
where
\begin{equation}
M=\left[1+\frac{m^2}{\alpha K}-\frac{\beta+\frac{m^2}{\alpha}}{4K}\Delta_y^T\Delta_y\right].
\label{67}
\end{equation}
Here we used the property
\begin{equation}
S^TS+\Delta_y^T\Delta_y=4.
\label{68}
\end{equation}

Next we integrate out the component $a_{1}$ in the action \eqref{66} to obtain
\begin{eqnarray}
\label{69}
S&=&\int dtdx\sum_y \left\{-\frac{1}{\pi}\partial_x\tilde{\theta}_{y+1/2}\partial_t\tilde{\varphi}_{y+1/2}+\frac{\sqrt{K}}{2\pi}\partial_x\Delta_y^T\tilde{\varphi}_{y+1/2}A_{0,y}-\frac{\gamma m^2}{8\pi\left(\alpha K+m^2\right)}\left(\partial_xS^T\tilde{\varphi}_{y+1/2}\right)^2\right\}\nonumber\\ \nonumber 
&+&\int dtdx\sum\limits_y\left\{-\frac{\gamma}{2\pi K^{2}\left(\alpha K+m^2\right)}\left(S^T\Delta_y^{-1T}\left(\partial_x\tilde{\theta}_{y+1/2}-\frac{\sqrt{K}}{2}\Delta_y A_{1,y}\right)\right)^2\right\}\\ \nonumber
&+&\int dtdx\sum\limits_y\left\{-\frac{m\gamma}{\pi K\left(\alpha K+m^2\right)}\left(\partial_x\tilde{\varphi}_{y+1/2}\right)S^T\Delta_y^{-1T}\left(\partial_x\tilde{\theta}_{y+1/2}-\frac{\sqrt{K}}{2}\Delta_y A_{1,y}\right)\right\}\\ \nonumber
&+&\int dtdx\sum\limits_y\left\{-\frac{\omega\alpha K^3+\gamma}{2\pi\alpha K^3}\left(\partial_x\tilde{\theta}_{y+1/2}-\frac{\sqrt{K}}{2}\Delta_y A_{1,y}\right)^2\right\}\\ \nonumber
&+&\int dtdx\sum\limits_y \left\{-\frac{\lambda\left(\alpha K+m^2\right)+\gamma m^2}{8\pi\left(\alpha K+m^2\right)}\left(\partial_x\Delta_y^T\tilde{\varphi}_{y+1/2}\right)^2 +g_{y,y+1} \cos \left(2\tilde{\theta}_{y+1/2}/\sqrt{K}-\epsilon_{\perp} A_{2,y}\right)\right\}\\ \nonumber
&+&\int dtdx\sum\limits_{y,y'} \frac{\gamma}{2\pi}\left(\partial_x\tilde{\varphi}_{y+1/2}-\frac{m}{\alpha K^{2}}S^T\Delta_y^{-1T}\left(\partial_x\tilde{\theta}_{y+1/2}-\frac{\sqrt{K}}{2}\Delta_y A_{1,y}\right)\right)\left(\Delta_y^TW\Delta_y\right)_{yy'}\times\\ 
&\times&\left(\partial_x\tilde{\varphi}_{y'+1/2}-\frac{m}{\alpha K^{2}}S^T\Delta_y^{-1T}\left(\partial_x\tilde{\theta}_{y'+1/2}-\frac{\sqrt{K}}{2}\Delta_y A_{1,y'}\right)\right),\nonumber\\
\end{eqnarray}
with $W$ implicitly defined by
\begin{equation}
M^{-1}=\frac{\alpha K}{\alpha K+m^2}+\Delta_y^TW\Delta_y\label{70},
\end{equation}
or directly via
\begin{equation}
W=\frac{\left(\beta\alpha+m^2\right)\alpha K}{4\left(\alpha K+m^2\right)^2}\left[1-\frac{\beta\alpha+m^2}{4\left(\alpha K+m^2\right)}\Delta_y^T\Delta_y\right]^{-1}.\label{71}
\end{equation}
The interaction mediated by the term $\Delta_y^TW\Delta_y$ decays exponentially with $\left|y-y'\right|$ and does not affect the universal long-distance properties of the model. Since, we are interested in the low-energy effective theory, we will ignore this term from now on. By comparing (\ref{69}) with (\ref{63}), we arrive at the following identifications between parameters:
\begin{equation}
\frac{\gamma}{K^2\left(\alpha K+m^2\right)}=\mathsf{v},~~~\omega+\frac{\gamma}{\alpha K^3}=\mathsf{u},~~~\text{and}~~~
\lambda=\mathsf{v}\left(1+k^2m^2\right).
\label{74}
\end{equation}
Therefore, these choices help us rewrite the nonlocal action \eqref{63} as the local model \eqref{64} with an emergent gauge field.


\bibliography{revised_prb_07_18_2022.bib}

\end{document}